\documentclass[preprint,aps,prl,superscriptaddress]{revtex4-2}

\usepackage{graphicx}
\usepackage{amsmath,amssymb}
\usepackage{siunitx}
\usepackage{hyperref}

\begin{document}

\preprint{APS/123-QED}

\title{Nonlinear response of soft hair beds to Poiseuille flows}

\author{M.S. Suryateja Jammalamadaka}
\affiliation{%
Massachusetts Institute of Technology, USA
}%
\author{Jonas Smucker}%
\affiliation{
 Center for Nonlinear Dynamics, University of Texas at Austin, USA
}%
\author{J. R. Alvarado}
\affiliation{
 Center for Nonlinear Dynamics, University of Texas at Austin, USA
}%

\date{\today}

\begin{abstract}
 Biological surfaces with micrometer-scale protrusions, such as microvilli, crustacean hairs, and cilia, often interact with pressure-driven fluid flow, resulting in a two-way elastoviscous problem. Characterizing their response to flow can enable applications in microfluidics, bioinspired engineering, and smart materials. Here, we investigate a biomimetic hair system subjected to pressure-driven flow experimentally and theoretically. We show that the rescaled resistance and rescaled pressure of various hair and chamber conditions collapse into an inverse power law after a critical dimensionless pressure, yielding one characteristic response across conditions. Our model predicts the behavior of angled hairs under Poiseuille flow along and against the grain, with the latter exhibiting significantly higher resistance. Finally, we demonstrate a conceptual application of angled hair beds to prevent backflow during intravenous therapy. This work establishes a unified model and experimental characterization of hair bed behavior in pressure-driven flows, advancing understanding of hair–flow interactions and laying the foundation for innovative applications.
\end{abstract}

\maketitle


\section{\label{sec:level1}Introduction}

Hair-like structures immersed in fluids are ubiquitous in nature and serve diverse functional roles. These hairy structures are known to increase surface area and aid in mass transport. They interact with fluids by absorbing or secreting biomolecules of interest to cells \cite{Li2023}, reducing shear forces to enhance surface slip \cite{Bauer2013}, and even driving fluid flows through synchronous active beating \cite{Kanale2022, Ling-2024-Nat.Phys.}. The tongues of many vertebrates are covered in papillae with sensory functions \cite{Harper2013}.  Blood vessels are coated in the glycocalyx, a layer of flexible and stiff polymer chains which facilitate mechanotransduction and protect cells from fluid shear forces \cite{Weinbaum2003, Weinbaum2007, Reitsma2007, VanTeeffelen2007, Mitsoulas-2022-Phys.Rev.Fluids}. The deformable carpet of microvilli of leukocytes play a key role in cell adhesion \cite{Wu2018}. By leveraging nature’s multifunctional hairy surfaces, bioinspired hairs and artificial cilia have emerged as versatile platforms for sensing, adhesion, and surface engineering enabling innovations \cite{lepora2018tacwhiskers, dong2020, cui2022}.

A hallmark feature of biological hair beds is their deformation in response to fluid flows, which influences their drag and their function \cite{Pramanik-2024-PhysicsofFluids}. Seminal work by Vogel demonstrated how tree-leaf reconfiguration results in the reduction of the drag coefficient as wind speed increases \cite{vogel1989}. Similar reconfiguration effects were found in underwater aquatic vegetation \cite{LuharNepf2011, Monti-2023-SciRep}. Parameters such as hair length, density, diameter, and physical composition affect the elastic properties of the hairs as well as hairs' interactions with fluid flows \cite{Seale2018}. In particular, the deformability of biological hairs affects their biological function \cite{Seale2018}. For example, bending of the supra-orbital whiskers plays an important role in detecting air flow in rats. \cite{Mugnaini2023}. Similarly, in the Venus flytrap (\textit{Dionaea muscipula}), bending of the trigger hairs initiates an action potential that initiates trap closure \cite{Yang2009}. Likewise, the stereocilia of inner ear hair cells deflect toward the kinocilium, thereby increasing tension on tip links, opening mechanically-gated ion channels, and thereby transducing graded receptor potentials \cite{Qiu2018}, \cite{Gillespie2009}. Together, these examples highlight the importance of deformability in biological fluid-structure interactions.

To understand the interplay between deformability and fluid flow effects, studies have built experimental model systems that mimic reconfiguration in natural systems. The advantage of model systems is the ability to closely relate to analytical \cite{smucker2022} and numerical models \cite{Rahimi-2022-PhysicsofFluids, Pang-2025-PhysicsofFluids, Sun-2025-PhysicsofFluids}. Experiments on flexible plates \cite{gosselin2010, deLangre2012} and hair beds \cite{Alvarado2017, Thomazo-2020-Phys.Rev.E, EtienneJambonPuillet2026} quantified drag reduction due to reconfiguration. At high velocities, turbulent flows can result in traveling waves \cite{brucker2012}. The resulting flutter increases drag, though potentially not as strongly as reconfiguration decreases drag \cite{Leclercq-2018-Proc.A}. Hair beds entrain air during immersion \cite{Nasto2016}, dissipate the kinetic energy of impact droplets \cite{nasto2019drop}, locally trap phases in interfacial flows \cite{ushay2023}, and govern tribological properties \cite{Peng-2021-ExpMecha}. A unified response framework for pressure‑driven flow through elastic hair beds, one that spans all packing fractions for straight arrays and is experimentally validated for angled arrays, remains lacking. Establishing such framework would enable engineering‑ready design and control.

This work establishes a unified model and experimental characterization of hair bed behavior in pressure-driven flows. We first derived the governing equations for deformable hairs in the Poiseuille flow and then simplified them into a model that captures their behavior. Consequently, a single plot of rescaled resistance versus dimensionless pressure unifies the response of straight hair beds across different geometries, revealing an inverse power law within a selective dimensionless pressure range, where the exponent depends on channel geometry. Furthermore, angled hairs exhibit a pronounced nonlinear response that strongly varies with the anchor angle. These insights not only advance our understanding of hair–flow interactions but also lay the foundation for innovative applications.

Building on these insights, we envision applications in biomedical engineering, microfluidics, and smart material design such as filtering impurities in fluidic pipes, trapping bacterial or viral particles in air purification systems, designing microfluidic diodes \cite{Brandenbourger-2020-Phys.Rev.Fluids}, memristors \cite{Martinez-Calvo-2024-NatCommun}, check valves for hypodermic needles, and self-cleaning surfaces.

\begin{figure}[htp] 
\centering \includegraphics[width=1\textwidth]{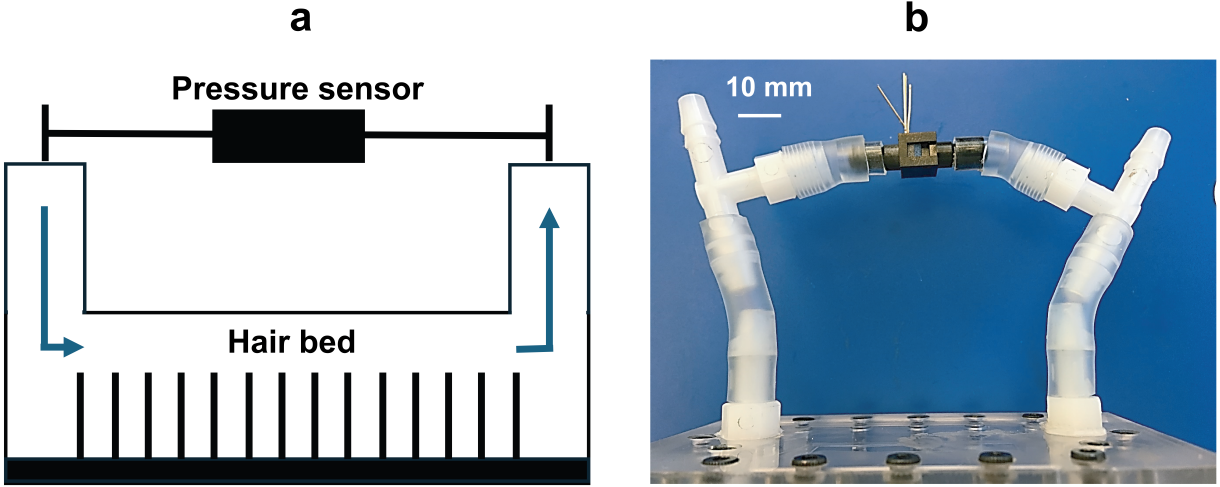} 
\caption{(a) Schematic of the experimental system. A hairbed placed in a flow chamber experiences a pressure differential between inlet and outlet, measured by a pressure sensor. (b) Front view of the fabricated experimental setup.}
\label{fig:fig1} \end{figure}

To investigate the behavior of hair beds under Poiseuille's flow, we fabricated straight and angled arrays with packing fractions $\varphi = 0.01, 0.03, 0.10,$ and $0.22$, channel heights $H = \SI{3.175}{\milli\meter}, \SI{4.5}{\milli\meter}, \SI{6.5}{\milli\meter}$, and hair radius $a=\SI{125}{\micro\meter}$, following Refs.~\cite{Nasto2016,Alvarado2017} (see SI for details). We imposed pressure-driven flow in a chamber instrumented with an inlet--outlet pressure sensor (Fig.~1a,b) and calculated the hydraulic resistance $R=\Delta p/Q$ for all geometries (Fig.~3a). We then developed a theoretical model and nondimensionalized resistance and pressure, revealing a distinct nonlinear dependence of $\tilde{\mathcal{R}}$ on $\Pi$ (Fig.~3b,c). Finally, we tested angled beds with anchoring angles $\theta_{0}=\pm\SI{10}{\degree}, \pm\SI{20}{\degree}, \pm\SI{30}{\degree}, \pm\SI{40}{\degree}$ at $\varphi=0.22$ and projected height $h=\SI{3.175}{\milli\meter}$, both along and against the grain, and compared with the model predictions (Fig.~4a). The resistance–pressure trends in Fig.~3a, which span packing fraction, channel height, and hair length, have an underlying organizing principle. To reveal it, we develop a minimal model of a hair‑coated channel that isolates the dominant geometric and material parameters and the relevant dimensionless groups.

\begin{figure}[htp] 
\centering \includegraphics[width=1\textwidth]{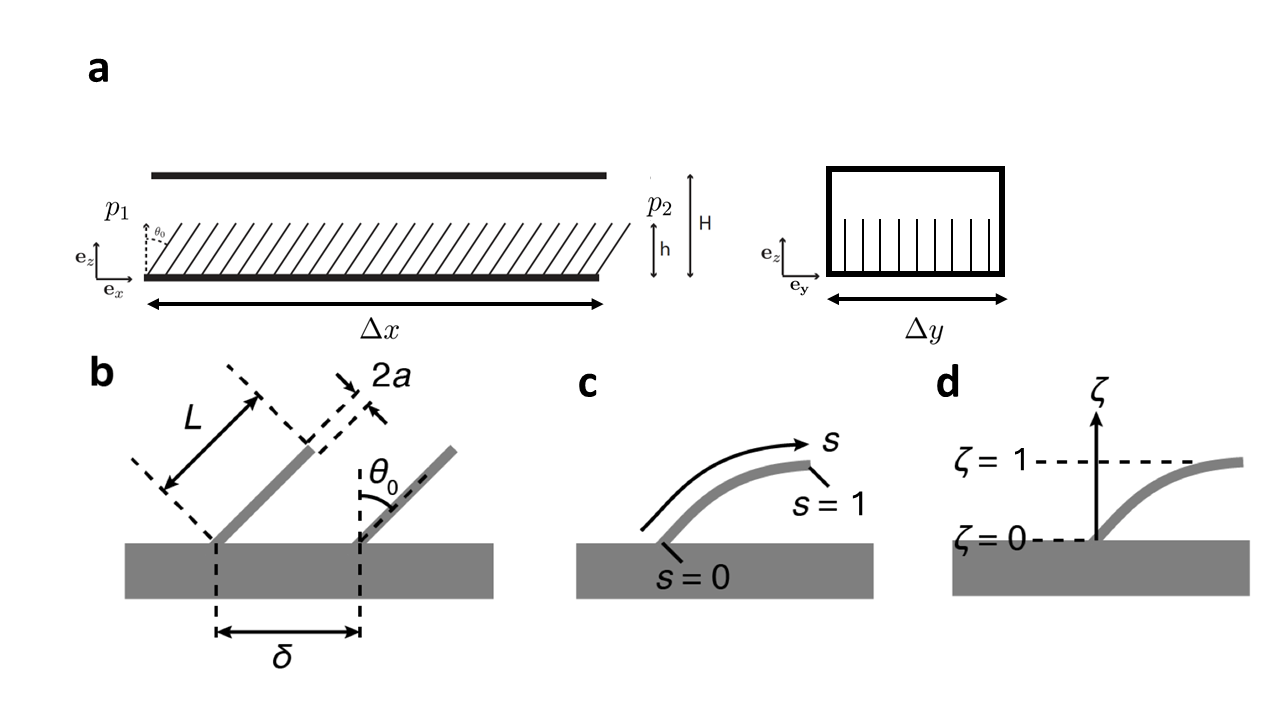} 
\caption{ (a)  Schematic of theoretical hair bed system in Poiseuille flow. A hair bed with anchoring angle \(\theta_0 \) and orthogonal height $h$, placed in flow chamber ( $\Delta x$ x $\Delta y$ x $\Delta z$) on one side with a pressure difference of $p_1$-$p_2$ between the inlet and outlet.  (b) Geometry of the hair bed system. Each hair has a length of $L$ with a radius of $a$, separated from each other at the anchoring surface by \(\delta \). (c, d) Coordinate system one to parameterize hair shape. $s$ is a normalized curvilinear coordinate that follows the centerline of bent hair and \(\zeta \) is the orthogonal coordinate from the anchoring surface. $s$ $\in$  [0, $1$] and \(\zeta \) $\in$  [0, 1]. }
\label{fig:fig2} \end{figure}

We consider the simplest geometry: two parallel planes with the lower plane coated with deformable hairs (Fig.~2a). The chamber has length $\Delta x$ in $x$ and width $\Delta y$ in $y$, with plate separation $H \equiv \Delta z$ measured at the hair bases (Fig.~2a). The hair tips reach height $h$ (projected along $z$) (Fig.~2a). Each hair has diameter $2a$, length $L$, elastic modulus $E$, and anchoring angle $\theta_{0}$ (Fig.~2b). The inter-hair spacing $\delta$ sets the packing fraction $\varphi$ (definition and lattice details in the SI). A pressure difference $\Delta p = p_{1}-p_{2}$ applied along $x$ drives fully developed flow that varies only with $z$,
$\mathbf{u}(\mathbf{x}) = u(z)$. When fluid flows past the bed, the hairs bend (Fig.~2c,d). To describe the deformed shapes, we use two coordinates: $\zeta$, the orthogonal distance from the anchoring surface along $z$ (Fig.~2c), and the curvilinear arclength coordinate $s$ along the hair centerline (Fig.~2d). Further geometric and modeling details are provided in the Supplemental Material.

We model flow within the hair bed using Darcy’s law,

\begin{equation}
u_{\mathrm{hairs}}(\zeta)
= -\,\frac{K\!\left(\theta(\zeta)\right)}{\eta}\,\nabla p
= \frac{K\!\left(\theta(\zeta)\right)}{\eta}\,\frac{p_{1}-p_{2}}{\Delta x},
\end{equation}

where \emph {K} [\SI{}{\square\meter}]  is the permeability, which depends on the local fiber orientation $\theta(\zeta)$ and the packing fraction $\varphi$. The orientation- and density-dependence of $K$ is constructed by interpolating between the sparse and dense limits reported in Ref.~\cite{Gopinath2011} for any fiber orientation $\theta \in [0,\frac{\pi}{2}]$ and fiber density $\varphi \in [0,1]$ (SI, Eqs.~(S3) and (S4)). In our implementation we include a prefactor $\varphi$ in these limiting forms and use the dimensionless permeability $ \kappa = \frac{K}{a^{2}}.$
Alternative predictions obtained directly from the permeability of Ref.~\cite{Gopinath2011} are provided in the Supplemental Material (Fig.~2 in SI).

Darcy's model describes the flow within the bulk of the hair bed but breaks down near the boundaries at the hair base and at the hair tips. At the hair base, the physically appropriate boundary condition is no-slip on the solid wall, which would generate a viscous boundary layer within the bed. For simplicity, we neglect this boundary layer and do not explicitly enforce no-slip at the base. At the interface between the hair bed and the overlying gap, one would generally impose continuity of both velocity and shear stress. In the present work, we assume that the shear stress in the tip layer is borne by the hair elasticity and therefore do not enforce shear-stress continuity across this interface, while retaining continuity of velocity. Finally, at the upper boundary (the hair-free plane), we impose a no-slip boundary condition.

Using these assumptions, we derive the flows through the hair bed and flow through the gap (SI, Eqs.~(S6) and (S8)). The total volumetric flow rate $Q$ is the sum of these contributions (SI, Eqs.~(S9) and (S10)). The hydraulic resistance of the system is then defined as the ratio of the pressure drop to the total flow rate, $ (p_{1} - p_{2})/Q$.

\begin{equation}
\label{eq:resistance_main}
 R =\frac{\eta\mathcal{L}}{\mathcal{W}H^3}\mathcal{R} = \frac{\eta \mathcal{L}}{\mathcal{W}H^3} \left[\hat{L}\hat{h}\overline{\kappa}\hat{a}^{2} + \frac{(1-\hat{L}\hat{h})^3}{12} + \frac{1}{2}\kappa^{+}\hat{a}^2(1-\hat{L}\hat{h}) \right]^{-1}
\end{equation}

with $\hat{h}=h/L$ the dimensionless hair-tip height, $\hat{L}=L/H$ the dimensionless hair length, and $\hat{a}=a/H$ the dimensionless hair radius. The resistance $\mathcal{R}$ is nondimensionalized by extracting the geometric prefactor $\eta \mathcal{L}/(2 \mathcal{W} H^{3})$, which depends only on the channel dimensions. The mean dimensionless permeability of the bed is $\overline{\kappa} = \frac{1}{h} \int_{0}^{h} \kappa\!\big(\theta(\zeta)\big)\,\mathrm{d}\zeta$, and the tip permeability is $\kappa^{+} = \kappa\!\big(\theta(\zeta=1)\big)$.

Note that $\hat{h}$, $\hat{L}$, $\bar{\kappa}$, and $\kappa^{+}$ are all approximately $\mathcal{O}(10^{-1}\text{--}10^{1})$ (SI Tables II-III). The parameter $\hat{a}$ controls the balance between the first term (arising from flow through the hairs) and the last two terms (arising from gap flow). 

So far we have remained agnostic about the shape of the hairs. We assume that all hairs deform identically, which allows us to characterize the shape of a single representative hair. The effective height $\hat{h}$ in \eqref{eq:resistance_main} is unknown and is given by
$\hat{h} = \int_{0}^{1} \cos\!\big(\theta(\hat{s})\big)\,\mathrm{d}\hat{s}$. We therefore solve for $\theta(s)$, the angle that the centerline makes at point \emph{s} with respect to the $z$-axis, which quantifies the hair shape. To obtain  $\theta(s)$, we first determine the shear stress at the hair tip and the moment acting on the hair due to the spatially varying drag force of the fluid. Enforcing a torque balance projected onto the $y$-axis then yields an integro--differential equation for  $\theta(s)$, which we solve to recover the hair shape.

We derive the shear stress at the hair tip, $\tau_{z=h}$, from the continuity of shear stress (SI, Eq.~(S11)). Using slender-body theory, we obtain the resulting bending moment  $\vec{M} (s)$ at the base due to hydrodynamic drag on the filament (SI, Eqs.~(S12) and (S13)). Projecting the torque balance onto the $y$-axis and combining $\tau_{z=h}$, $\vec{M} (s)$, and the internal elastic torque yields the following equilibrium equation:

  \begin{equation}
   \label{eq:governingeq}
  \begin{split}
    0 = \frac{\textrm{d}^2\theta(s)}{\textrm{d} s^2} - \Pi \textrm{cos}\theta (s) \Big( \frac{\hat{a}^2\kappa^{+}}{1-\hat{L}\int_0^1 \textrm{cos}\theta(\hat{s})d\hat{s}}  - \frac{1}{2}\Big(1- \hat{L}\int_0^1 \textrm{cos}\theta(\hat{s}) d\hat{s}\Big) 
    \\
    + \frac{4 \varphi \hat{L}}{log(\frac{L}{a})}\int_s^1 \overline{\kappa}( 1- \frac{\sin^2\theta (\hat{s})}{2})
   {\textrm{d}\hat{s}} \Big) - \frac{\varphi \hat{L}} {log(\frac{L}   {a})} \Pi \sin\theta (s) \int_s^1 \overline{\kappa} \sin2\theta(\hat{s}) {\textrm{d}\hat{s}}
    \end{split}
 \end{equation} 
\begin{equation}
\label{eq:bigpi}
     \Pi= \frac{\pi a^2 L^2 H \Delta p}{EI\varphi\mathcal{L}} = \frac{4HL^2\Delta p}{Ea^2\mathcal{L}\varphi}
 \end{equation}

 The equation \eqref{eq:governingeq} reveals  a dimensionless group \(\Pi \) that scales with the pressure difference \(\Delta p \) \eqref{eq:bigpi}.  \(\Pi \) measures the balance between hydrodynamic and elastic forces acting during hair deformation (SI, Eq.~(S15)). The equation of equilibrium \eqref{eq:governingeq}simplifies once we neglect terms proportional to the small parameter  $ \frac{\varphi \hat{L}} {\log\frac{L} {a}}$, which is justified in the slender‑rod limit (SI, Eq.~(S16)):

  \begin{equation}
  \label{eq:governingeq_simplified}
    0 = \frac{\textrm{d}^2\theta(s)}{\textrm{d} s^2} - \Pi \cos\theta (s) \left( \frac{\hat{a}^2\kappa^{+}}{1-\hat{L}\int_0^1 \cos\theta(\hat{s})\textrm{d}\hat{s}}  - \frac{1}{2}\left(1- \hat{L}\int_0^1 \cos\theta(\hat{s})\textrm{d}\hat{s} \right)\right)
 \end{equation}

 (6) is a differo-integral equation of equilibrium with the following boundary conditions: $\theta(s=0)=\theta_0 \; \textrm{and} \; \theta'(s=1)=0$

To compute the dimensionless resistance $\mathcal{R}$, we require five parameters: the anchoring angle $\theta_{0}$, the dimensionless channel height $\hat{L}$, the dimensionless pressure $\Pi$, the packing fraction $\varphi$, and the dimensionless hair radius $\hat{a}$. This set exceeds the parameter set for the shear--driven case~\cite{Alvarado2017}, where only three quantities determine the impedance: $\theta_{0}$, $\hat{L}$, and the driving velocity $\widetilde{v}$ (analogous to $\Pi$). The two additional parameters, $\varphi$ and $\hat{a}$, control the additional flow through the hair bed that arises under pressure--driven flow and is absent in the shear--driven configuration. Given these inputs, we compute $\theta(s)$ from \eqref{eq:governingeq_simplified} using an iterative scheme initialized with guesses for $\hat{h}$, $\bar{\kappa}$, and $\kappa^{+}$, updating until convergence to a physically feasible solution.
   
To build intuition for the resistance $\mathcal{R}$, we derive the limiting values $\mathcal{R}_{0}$ and $\mathcal{R}_{\infty}$ corresponding to $\Pi \to 0$ and $\Pi \to \infty$, respectively (see SI for definitions and derivations, Eq.~(S18)). We then test the theoretical assumptions by comparing model predictions with experiments. Using the measured pressure drops, flow rates, and resulting resistances, we compute the dimensionless pressure $\Pi$ and the rescaled resistance $\tilde{\mathcal{R}}$ for straight hair beds with packing fractions $\varphi = 0.22, 0.10, 0.03$, and $0.01$ and dimensionless channel heights $\hat{L} = 0.78, 0.83$, and $0.87$ (see SI for justification and the precise normalization,  Eq.~(S19)). The Reynolds number for these experiments spans the ranges $Re_{\text{gap}} \in [1.3\times10^{-6},\, 4.12\times10^{1}]$ and $Re_{\text{hairs}} \in [6.9\times10^{-7},\, 7.54\times10^{-3}]$ across all configurations (SI Table IV).

Except for the extremely low packing-fraction case ($\varphi \approx 0.01$), the experimental results are qualitatively consistent with the computational model and quantitatively within an order of magnitude for $\varphi = 0.22$, $0.10$, and $0.03$ (Fig.~3; error factors listed in SI Table~I). We define the lowest packing fraction as $\varphi \lesssim 0.01 \approx 0.05\,\varphi_{\max}$, with $\varphi_{\max}=0.22$ (i.e., 5\% of the maximum explored value). For experiments with $\varphi \lesssim 0.01$, the model reaches its lower validity bound and fails to reproduce the measurements (SI Fig.~1). We anticipate this discrepancy arises because fluid--structure interaction becomes minimal and/or Darcy's law no longer holds at such low packing fractions. Darcy's law assumes a homogeneous porous medium; however, at extremely low $\varphi$ the permeability can be very large and the flow field heterogeneous, with distinct near-hair and interstitial free-flow regions and possible secondary flows, as reported for rigid hair arrays~\cite{Hood2019}. Such effects are expected to decrease $\tilde{\mathcal{R}}$ relative to the present model's predictions, and we indeed observe this behavior at $\varphi = 0.01$. This heterogeneity lies outside the model's assumptions and is beyond the scope of the present work.

Across all examined values of $\varphi$ and $\hat{a}$ (Fig.~3b), $\tilde{\mathcal{R}}$ and $\Pi$ obey an inverse power-law relation for $\Pi > 5$. Remarkably, for fixed channel geometry ($\hat{L}$ constant), the model predictions collapse onto a single curve spanning $0.05 < \tilde{\mathcal{R}} < 1$, indicating that the $\tilde{\mathcal{R}}$--$\Pi$ dependence becomes effectively independent of the specific hair-bed geometry. When the channel geometry is varied, as in the cases $\hat{L} = 0.59$ and $0.46$, the curves shift systematically: decreasing $\hat{L}$ reduces the sensitivity of $\tilde{\mathcal{R}}$ to changes in $\Pi$ (Fig.~3b).

Our experimental measurements display the same behavior. For $\Pi > 5$, the rescaled resistance $\tilde{\mathcal{R}}$ collapses cleanly onto the predicted inverse power law for $\varphi = 0.1$ and $0.22$, independent of $\hat{a}$ (Fig.~3c). At lower packing fractions, our measurements do not reach the $\Pi > 5$ regime and therefore do not access the collapse.

The power-law exponent in the inverse relation between $\tilde{\mathcal{R}}$ and $\Pi$ depends on the dimensionless length $\hat{L}$ \eqref{eq:powerlaw}. As $\hat{L}$ increases, the sensitivity of $\tilde{\mathcal{R}}$ to $\Pi$ increases. For $\varphi = 0.1$, the exponents are $\alpha = -1.28$, $-0.89$, and $-0.57$ for $\hat{L} = 0.78$, $0.59$, and $0.46$, respectively. Moreover, $\alpha$ depends systematically on $\hat{L}$ as shown in \eqref{eq:powerlaw} with $\beta = 0.482$ (approximately $0.5$) and $R^{2} = 0.993858$.

\begin{equation}
\label{eq:powerlaw}
    \alpha = -0.7  
\left[ \frac{\hat{L}}{1-\hat{L}}\right]^{\beta}
\end{equation}

Taken together, these results show that the high-dimensional parameter space spanned by $\hat{a}$, $\varphi$, bending, and their coupling to the flow reduces, in the appropriate regime, to a single dominant relation between $\tilde{\mathcal{R}}$ and $\Pi$. This collapse provides a simple predictive framework for estimating flow resistance across a broad range of hair-bed geometries, revealing a surprisingly simple organizing principle in an otherwise complex fluid--structure system.

We also solved \eqref{eq:governingeq_simplified} using the permeability formulation of Ref.~\cite{Gopinath2011}. Unlike the collapse presented in the main text, the permeability-based model of Ref.~\cite{Gopinath2011} does not collapse onto a single inverse power law. Instead, the results cluster according to packing fraction (see SI Fig.~2). For this reason, we report this result only in the Supplementary Information and refrain from using it in the main text.

\begin{figure}[htp] 
\centering \includegraphics[width=1\textwidth]{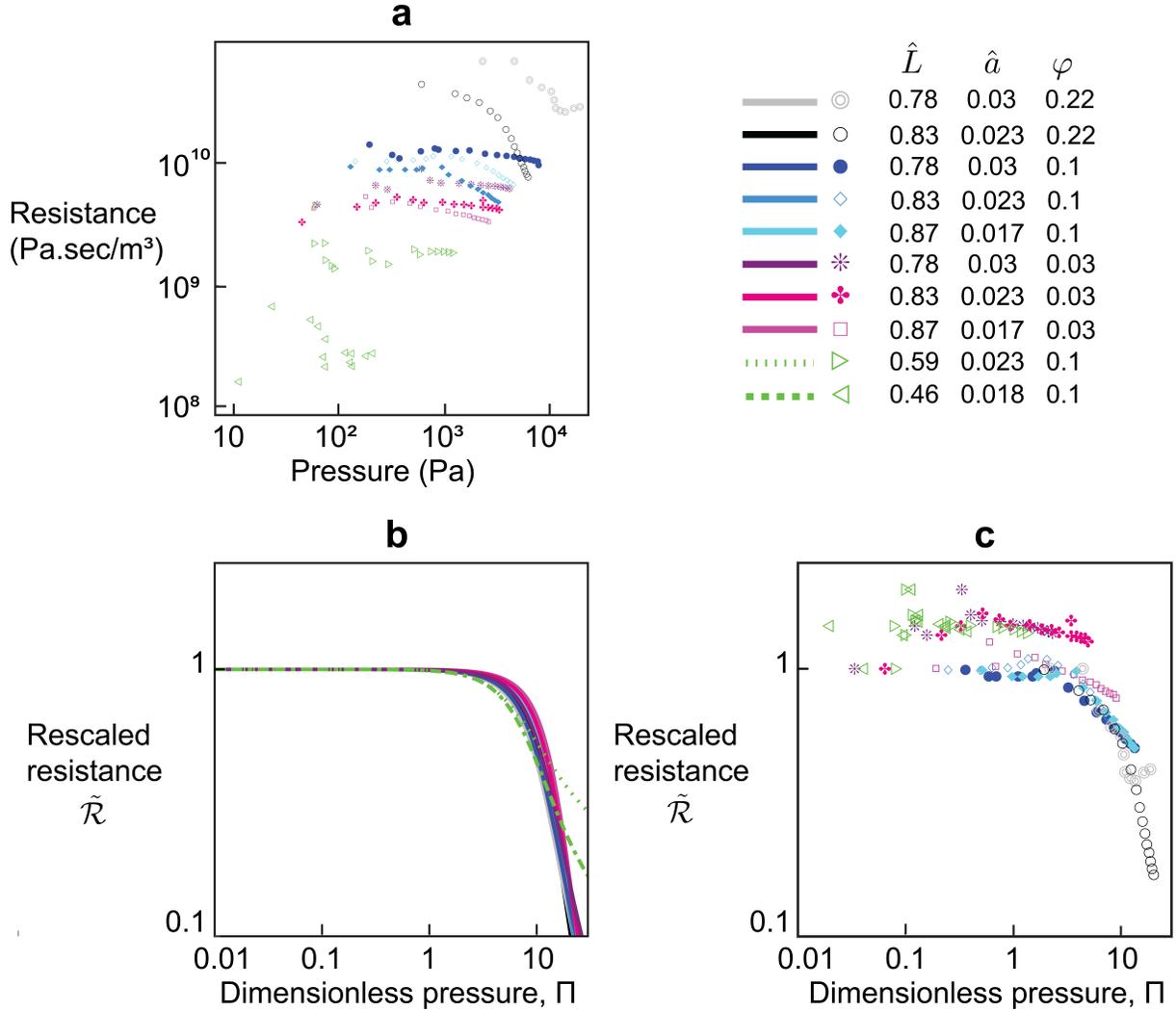} 
\caption{(a) Experimental outcomes of ten geometries of straight hair beds ($\hat{L}, \hat{a}, \varphi$) (b,c) Rescaled resistance ($\tilde{\mathcal{R}}$) as a function of dimensionless pressure $\Pi$ for ten geometries of straight hair beds ($\hat{L}, \hat{a}, \varphi$). Solid lines and dotted lines correspond to theoretical predictions for respective  hairbeds ($\hat{L}, \hat{a}, \varphi$). Symbols correspond to experimental outcomes for  respective hairbeds ($\hat{L}, \hat{a}, \varphi$).} 
\label{fig:Fig3} \end{figure}

We next tested whether the model captures the behavior of angled hairs oriented both along and against the grain. The experimental results show good qualitative and quantitative agreement with the model predictions. For hairs with positive $\theta$ (aligned with the grain), the normalized resistance $\tilde{\mathcal{R}}$ decreases nonlinearly with $\Pi$, similar to the straight--hair case but with the onset of nonlinearity occurring at smaller values of $\Pi$ (Fig.~4a). In contrast, for hairs oriented against the grain (negative $\theta$), $\tilde{\mathcal{R}}$ initially increases, reaches a peak, and subsequently decreases as $\Pi$ increases (Fig.~4a). This nonlinear trend mirrors the behavior observed in angled hair beds under Couette flow~\cite{Alvarado2017}. For $\theta = -40^{\circ}$, the hair bed contacts the ceiling, driving the inter--hair gaps to zero and causing numerical instability in the computational model. We therefore report its behavior only for $\Pi < 10$ (Fig.~4a). Increasing the magnitude of $\theta$ (more strongly against the grain) systematically elevates $\tilde{\mathcal{R}}$ at a fixed $\Pi$, indicating that larger tilt angles produce a more pronounced nonlinear response to small variations in $\Pi$ (Fig.~4a).

\begin{figure}[htp] 
\centering \includegraphics[width=1\textwidth]{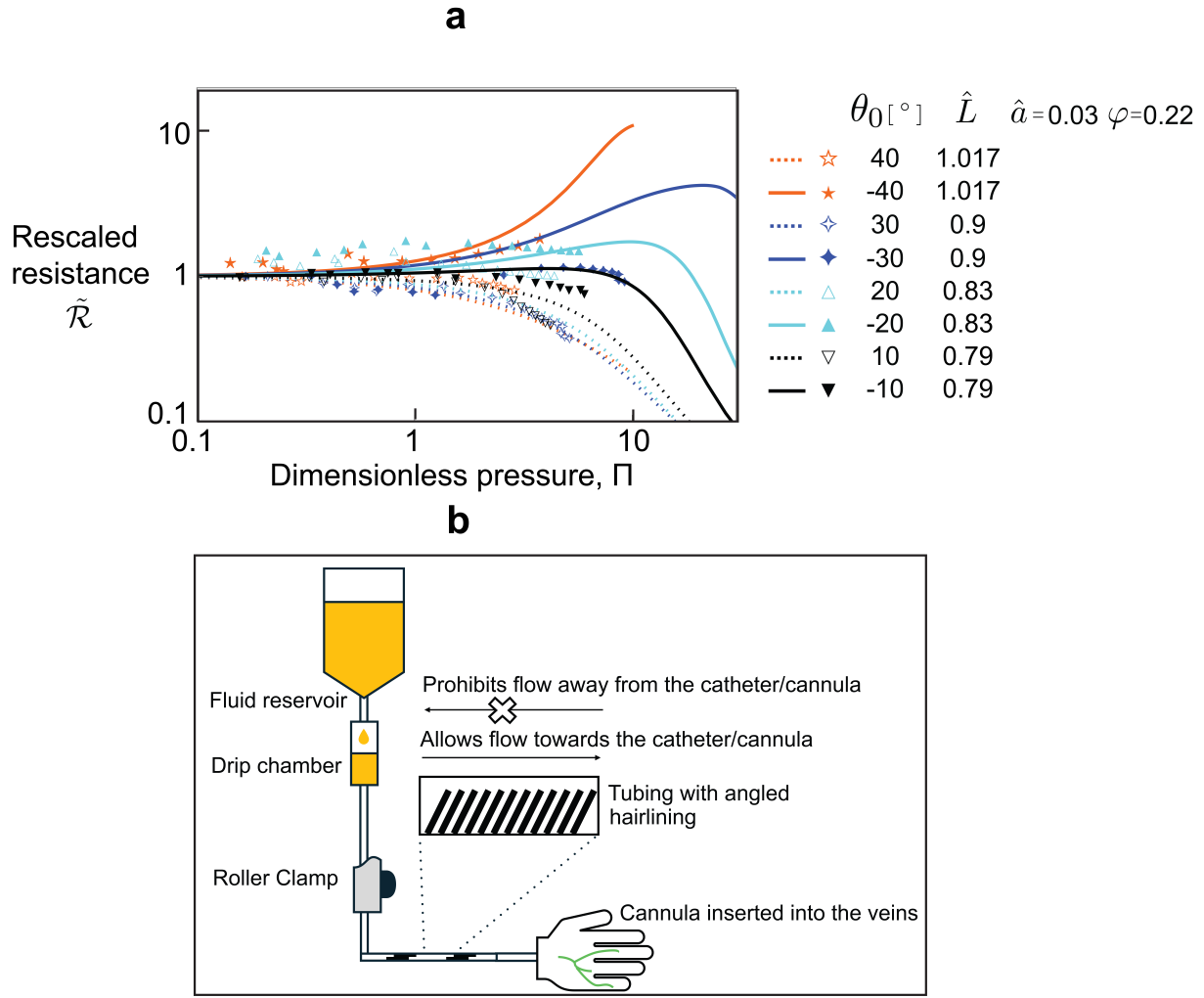} 
\caption{(a) Rescaled resistance ($\tilde{\mathcal{R}}$) as a function of dimensionless pressure $\Pi$ for eight geometries of angled hair beds ($\theta_{0}, \hat{L}, \hat{a}, \varphi$ ).  Solid lines and dotted lines correspond to theoretical predictions for aginst the grain and along the grain respectively. Symbols correspond to experimental outcomes for  respective hairbeds ($\theta_{0}, \hat{L}, \hat{a}, \varphi$). (b) Potential technological application of angled hair beds: Angled hair beds can prevent backflows in Intravenous Therapy by offering increased resistance. } 
\label{fig:Fig4} \end{figure}

Bioinspired hair structures have been exploited across a wide range of applications, including flow sensing~\cite{zhang2025bioinspired}, adhesive and friction--modulating surfaces~\cite{dong2020}, acoustic sensing~\cite{moshizi2024}, energy harvesting~\cite{khan2018}, microfluidic propulsion and mixing~\cite{sahadevan2022}, anti--fouling coatings~\cite{zhang2020}, self--cleaning surfaces~\cite{cui2021}, and soft robotics~\cite{gu2020}. The nonlinear response of angled hair beds demonstrated here suggests a new class of passive flow--control elements. One potential application is the prevention of backflow in intravenous (IV) therapy, where maintaining a stable pressure differential between the drip chamber and the venous pressure is critical~\cite{moorthy2020}. When the hydrostatic pressure in the drip chamber falls below a threshold, venous pressure can drive reverse flow, posing risks of blood loss and infection. Current mitigation strategies rely on frequent manual monitoring or sensor--based devices, which can be costly and inaccessible in resource--limited settings~\cite{moorthy2020}. A passive check valve integrated directly into the IV tubing provides a simple alternative (Fig.~4b). By lining a short section of the tubing with angled, deformable hairs, forward flow encounters low resistance (with the grain), while reverse flow experiences substantially greater resistance (against the grain) (Fig.~4b). Such a geometry--driven, passive mechanism could offer an effective and low--cost solution to backflow in IV systems.

In conclusion, we have developed a minimal model that captures the behavior of deformable hair beds in pressure--driven flow. In contrast to shear--driven configurations, the resistance in pressure--driven flow depends not only on the anchoring angle $\theta_{0}$, the dimensionless hair length $\hat{L}$, and the dimensionless pressure $\Pi$, but also on two additional geometric parameters: the packing fraction $\varphi$ and the dimensionless hair radius $\hat{a}$. The parameter $\hat{a}$ determines the balance between resistance of the hair bed and resistance of the gap. For straight hair beds across a range of packing fractions, the rescaled resistance follows an inverse power--law relationship with $\Pi$, with the exponent governed by dimensionless hair length $\hat{L}$. The model, however, breaks down at the lowest packing fractions ($\varphi \le 0.01$) where the Darcy law fails or due to minimal fluid structure interaction. Finally, we show that angled hairs exhibit a strongly nonlinear dependence of the dimensionless resistance on $\Pi$, with the nonlinearity becoming increasingly pronounced as the anchor angle increases. This tunable response may be exploited for controlling backflow in intravenous therapy and may offer new design principles for biomedical devices, smart materials, and microfluidic systems.

\newpage

\begin{acknowledgments}
We gratefully acknowledge Dr. Jean Comtet for providing support and guidance during the development of the initial modeling framework. We also thank Prof. A. E. Hosoi for generously granting access to her laboratory space, which enabled the experimental component of this study. We acknowledge helpful discussions with Prof. Anoop Rajappan.
\end{acknowledgments}
 
\bibliography{references_main}
\end{document}